\documentclass[a4paper]{jpconf}
\usepackage[T1]{fontenc}
\usepackage{ae,aecompl}
\usepackage{graphics,graphicx,dcolumn,bm,fleqn,epic,eepic,float}
\usepackage{amssymb,amsmath,multirow,rotate,color,float,times}
\usepackage{color}
\usepackage{soul}                    
\definecolor{red}{rgb}{1,0,0}
\definecolor{green}{rgb}{0,1,0}
\definecolor{blue}{rgb}{0,0,1}

\newcommand{\bite}{\begin{itemize}}
\newcommand{\eite}{\end{itemize}}
\newcommand{\benu}{\begin{enumerate}}
\newcommand{\eenu}{\end{enumerate}}
\newcommand{\bverb}{\begin{verbatim}}
\newcommand{\everb}{\end{verbatim}}

\newcommand{\beq}{\begin{equation}}
\newcommand{\eeq}{\end{equation}}
\newcommand{\barr}{\begin{array}}
\newcommand{\earr}{\end{array}}

\begin{document}

\title{Evaluating strong measurement noise in data series with simulated
annealing method}

\author{J.~Carvalho$^1$,F.~Raischel$^2$,M.~Haase$^3$,P.~Lind$^{1,2}$}

\address{$^1$Departamento de F\'{\i}sica, Faculdade de Ci\^encias  da Universidade de Lisboa, 1649-003 Lisboa, Portugal}
\address{$^2$Center for Theoretical and Computational Physics,
         University of Lisbon, 1649-003 Lisbon, Portugal}
\address{$^3$Institute for High Performance Computing, University of Stuttgart,  D-70569 Stuttgart, Germany}
\ead{raischel@cii.fc.ul.pt}

\begin{abstract}
Many stochastic time series can be described by a Langevin equation 
composed of  a deterministic and a stochastic dynamical part. 
Such a stochastic process can be reconstructed by means of a 
recently introduced nonparametric method, thus increasing the 
predictability, i.e.~knowledge of the macroscopic drift and the microscopic 
diffusion functions. 
If the measurement of a stochastic process is affected by additional strong 
measurement noise, the reconstruction process cannot be applied. 
Here, we present a method for the reconstruction of stochastic processes 
in the presence of strong measurement noise, based on a suitably parametrized  
ansatz. 
At the core of the process is the minimization of  the functional distance 
between terms containing the conditional moments taken from measurement data,
and the corresponding ansatz functions. It is shown that a minimization of 
the distance by means of a simulated annealing procedure yields better  results than a previously used Levenberg-Marquardt algorithm, which permits a rapid and reliable reconstruction of the  stochastic process. 
\end{abstract}


\section{Introduction}
\label{sec:int}

Physical systems often can be resolved on two different time scales, 
observing slowly varying macroscopic motion while  much faster 
microscopic interactions are perceived as an effective noisy driving. An adequate 
description of these systems can be obtained by a Langevin equation, 
which yields a deterministic drift term and a stochastic diffusion term. 
With respect to this description, reconstructing the drift and diffusion terms of an 
unknown process in time or scale from a data set  enables one to 
enhance the predictability of that process.
Recently, a nonparametric method of reconstruction based solely in the 
measured time series has been suggested \cite{friedrich97}.
The method  has been  applied successfully to data sets from  
fields such diverse as turbulent fluid dynamics \cite{nawroth07}, human 
movement \cite{vanMourik2006}, financial data \cite{friedrich00}, climate 
indices \cite{collette04,lind05}, and to electroencephalographic recordings 
from epilepsy patients \cite{prusseit08,lehnertz09}. For the case of finite time
steps \cite{ragwitzCorr2001,friedrichComm2002}, which can occur when the 
data is not available at sufficiently high sampling rates, recent 
improvements have been suggested \cite{kleinhans05,gottschall08,Anteneodo2009a}.

However, any experimental setup will record  additional measurement noise 
when recording a data series, which cannot be easily distinguished from the 
intrinsic  dynamical noise. 
The presence of measurement noise can destroy the Markovian properties of 
the data and lead to incorrect estimates for the drift and diffusion 
terms \cite{Lind2010a, kleinhans07}.

It has been suggested to separate the measurement noise from the dynamics of 
the measured variable using different predictor models or schemes for noise 
reduction \cite{schreiberbook,abarbanel93}.
Recently, another approach has been  suggested,  which allows reconstructing
the properties of stochastic time series affected by exponentially 
correlated Gaussian noise, using algebraic properties of the normal 
distribution \cite{lehle2010}.
Here, we revisit an approach presented in Refs.~\cite{boettcher06,Lind2010a}, 
which minimizes the functional distance between characteristic functions extracted 
from the measurement data and corresponding terms calculated from a 
suitably parametrized ansatz for the drift and diffusion functions.  
We show that using a Simulated Annealing 
(SA) \cite{saref} method for the optimization  
yields  results that are superior to the ones obtained  previously by means of the Levenberg-Marquardt 
(LM) method.

We start in Sec.~\ref{sec:framework} by describing the general framework
for extracting the evolution equation of a signal spoiled by strong
measurement noise, together with the amplitude of said measurement noise.
In Sec.~\ref{sec:optimization} both LM and SA methods are addressed and 
compared.
Section \ref{sec:conc} concludes the paper.

\section{Langevin approach for data sets with measurement noise}
\label{sec:framework}

We consider a one-dimensional Langevin process $x(t)$ defined as
\begin{equation}
\frac{dx}{dt} = D_1(x) + \sqrt{D_2(x)}\Gamma_t ,
\label{xlangevin}
\end{equation}
where $\Gamma_t$ represents a Gaussian $\delta$-correlated 
white noise $\langle\Gamma_t\rangle = 0$ and
$\langle\Gamma_t\Gamma_{t^{\prime}}\rangle = 2\delta(t-t^{\prime})$.
The Kramers-Moyal (KM) functions $D_1(x)$ and $D_2(x)$ are defined as 
\begin{eqnarray}
D_n(x) = \frac{1}{n!}\lim_{\tau \to 0}\frac{1}{\tau}{M}_n(x,\tau)
\label{KramersMoyal}
\end{eqnarray}
for $n=1,2$, where $D_1$ describes the deterministic drift and $D_2$ corresponds to a 
diffusion process. 
An estimate of the $n$-th order conditional moments of the data   
${M}_n(x,\tau)$  can be extracted directly from the measured time series as
\begin{eqnarray}
  \hat{M}_n(x_i,\tau) &=& \langle (x(t+\tau)-x(t))^n \rangle |_{x(t)=x_i} \, ,
  \label{gencondmom_x}  
\end{eqnarray}
the hat indicating that they are calculated from the measured data $x(t)$ directly, whereas the  brackets $<>$ denote averaging over suitable parts of the time 
series -- or ensemble averages, if available -- as described in 
Refs.~\cite{friedrich97,friedrich00,lind05,boettcher06}.
Thus, a nonparametric reconstruction of the stochastic process can be achieved.

In the following, we will consider a time series generated by integrating 
Eq.~(\ref{xlangevin}) with drift and diffusion coefficient assumed to be 
linear and quadratic forms respectively,
\begin{subequations}
\begin{eqnarray}
D_1(x) &=& d_{10} + d_{11}x \label{D1}\\
       & & \cr
D_2(x) &=& d_{20} + d_{21}x + d_{22}x^2 . \label{D2}
\end{eqnarray}\label{DD}\end{subequations}
Though we concentrate on the particular expressions for $D_1$ and $D_2$
given above, it should be stressed that they comprehend a large collection 
of different processes, such as Ornstein-Uhlenbeck processes \cite{boettcher06}.
The parameter $d_{10}$ can  always be eliminated through a simple
transformation $x\to x^{\prime}=x+d_{10}/d_{11}$. 
For the numerical examples in the remainder of the text, we will use 
the parameters $d_{10}=1, d_{11}=-1,  d_{20}=1,d_{21}=-1,d_{22}=1$.
\begin{figure}[htb]
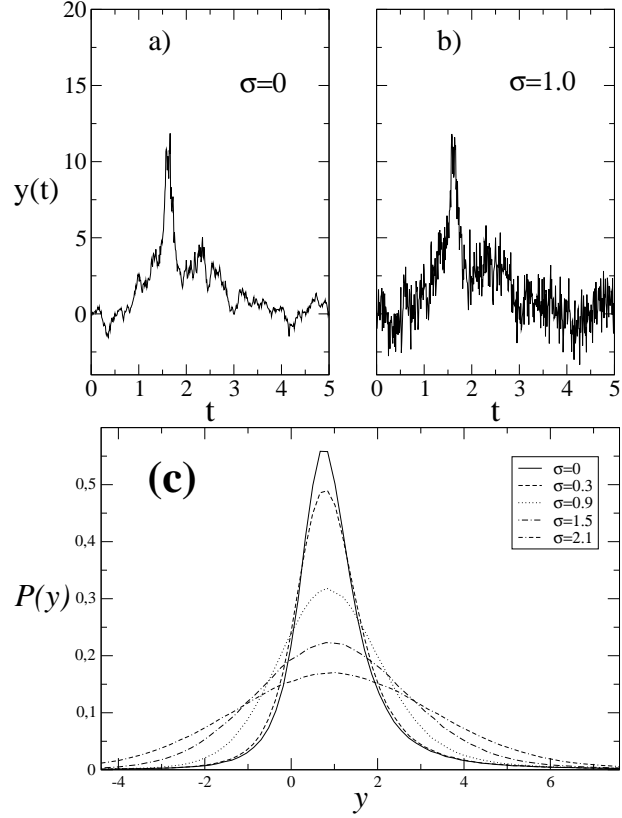

\begin{center}
\includegraphics*[width=8.0cm,angle=0]{fig01_joao.eps} 
\includegraphics*[width=8.0cm,angle=0,]{fig01c_joao.eps}
\end{center}
\caption{\protect Langevin time series with different measurement
  noise strengths. 
  Here we show a time series 
  {\bf (a)} without measurement noise, and 
  {\bf (b)} with strong measurement noise. 
  In {\bf (c)}  the probability density function $P(y)$ of the series
  with measurement noise (see Eq.~(\ref{yy})) is shown. 
  In all cases, the assumed time series $x(t)$ without measurement
  noise uses the coefficients $D_1(x)=1-x$ and $D_2(x)=1-x+x^2$,
  when integrating Eq.~(\ref{xlangevin}).}
\label{fig1}
\end{figure}

We now  consider the case that the signal of $x(t)$ is affected  by a 
Gaussian $\delta$-correlated measurement white noise, which leads to the 
series of observations
\begin{equation}
y(t) = x(t) + \sigma \zeta(t)
\label{yy}
\end{equation}
where $\sigma$ denotes the amplitude of the measurement noise. 
As Figs.~\ref{fig1}a and \ref{fig1}b show, although the trend of $x(t)$ 
is still visible in the presence of measurement noise, it is intuitively 
difficult to distinguish between the intrinsic dynamic noise and the 
measurement noise. 
Likewise, Fig.~\ref{fig1}c illustrates that  for increasing $\sigma$ one 
obtains broader probability density functions $P(y)$: the standard deviation 
of these distribution increases with the measurement noise, whereas the mean 
value remains constant.

In the presence of measurement noise, the conditional moments yield
non-zero finite values at $\tau=0$ and thus the limit 
$\lim\limits_{\tau\to 0} \frac{1}{\tau}\hat{M}_n(x,\sigma\ne 0,\tau)$ does not 
exist \cite{Lind2010a}. Likewise, the requirements for Markovianity of the measured 
time series $y(t)$ may not be satisfied \cite{kleinhans07}.

\begin{figure}[htb]
\begin{center}
\includegraphics*[width=8.5cm,angle=0]{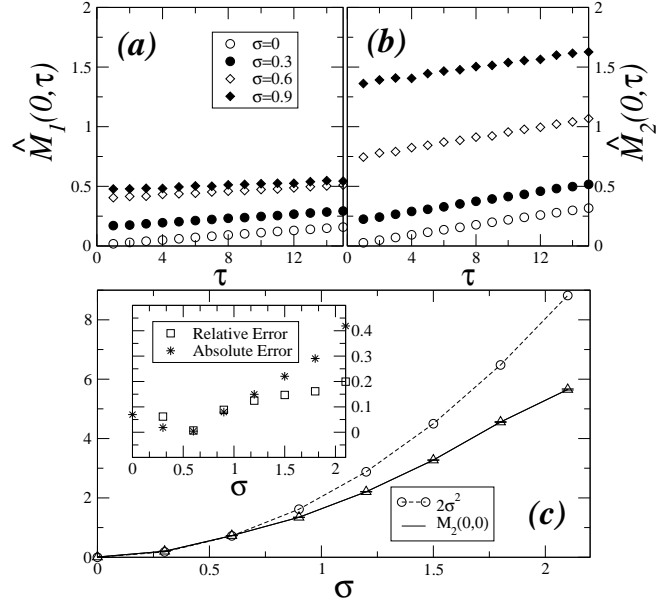}
\end{center}
\caption{\protect   
   Conditional moments 
   {\bf (a)} ${\hat M}_1(y_i,\tau)$ and 
   {\bf (b)} ${\hat M}_2(y_i,\tau)$ as a function of $\tau$, for
   bin $y_i=0$ and different measurement noise strengths $\sigma$. 
   The same $x(t)$ as in Fig.~\ref{fig1} was used.
   {\bf (c)} First estimate of the measurement noise (see text).}
\label{fig2}
\end{figure}

However, the two conditional 
moments \cite{friedrich97,friedrich00,lind05,boettcher06,Lind2010a}
$\hat{M}_n(y_i,\tau)$ ($n=1,2$) can be derived taking $y(t)$ instead
of $x(t)$ in Eq.~(\ref{gencondmom_x}).
Figure \ref{fig2} shows  plots of both conditional moments as a 
function of $\tau$ for different strengths of the measurement noise.
The linear dependence of $M_1$ and $M_2$ on $\tau$ makes it still possible to obtain  an estimate for  the `spoiled' KM-coefficients:
\begin{equation}
\hat{D_n}(y) = \frac{\hat{M}_n(y,\tau_2)- \hat{M}_n(y,\tau_1)}
                    {n!(\tau_2-\tau_1)} \, .
\label{Dprime}
\end{equation}

Within this description, an estimate for the amplitude of the measurement noise
yields $\sigma \approx \sqrt{\frac{\hat{M}_2(\mu,0)}{2}}$ 
(see Ref.~\cite{boettcher06,siefert03}), where $\mu$ is the average value of 
$y(t)$ data points in the time series. 
As shown in Fig.~\ref{fig2}c, such an estimate is not valid for sufficiently 
strong measurement noise, i.~e.~$\sigma\gtrsim 0.5$.

\begin{figure}[htb]
\begin{center}
\includegraphics*[width=8.0cm,angle=0]{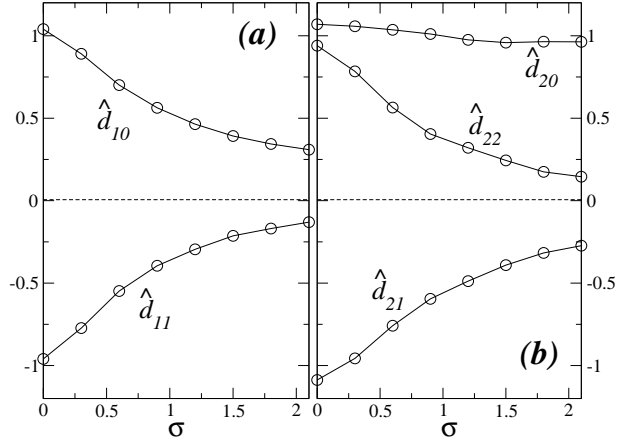}
\end{center}
\caption{\protect 
  Noise dependence of functions $\hat{D}_1(y)$ and $\hat{D}_2(y)$ 
  (see text and Eq.~(\ref{Dprime})).
  The underlying Langevin time series $x(t)$ without noise 
  is the same as in Fig.~\ref{fig1}.}
\label{fig3}
\end{figure}

The coefficients $D_1$ and $D_2$ cannot be  correctly estimated, though.
Figure \ref{fig3} shows how the estimated parameters 
$\hat{d}_{ij}$ deviate from the `true'  values $d_{ij}$ of the process given by 
Eq.~(\ref{DD}) as measurement noise increases. 
Notice that for $\sigma=0$ (left vertical axis in each plot of 
Fig.~\ref{fig3}) the estimated parameter values are approximately 
correct.

To correctly derive the drift and diffusion coefficients $D_1(x)$ and
$D_2(x)$ when $\sigma$ is strong, we consider the measured conditional
moments $\hat{M}_1(y_i,\tau)$ and $\hat{M}_2(y_i,\tau)$.
Since these conditional moments depend in a non-trivial way on both 
time $\tau$ and amplitude $y_i$, we approximate them up to first 
order on $\tau$:
\begin{subequations}
\begin{eqnarray}
\hat{M}_1(y_i,\tau) &=& \langle y(t+\tau)-y(t) \rangle |_{y(t)=y_i}= \tau \hat{m}_1(y_i) + \hat{\gamma}_1(y_i) + {\cal O}(\tau^2),            \label{yM1_2}\\
\hat{M}_2(y_i,\tau) &=& \langle (y(t+\tau)-y(t))^2 \rangle |_{y(t)=y_i}  = \tau \hat{m}_2(y_i) + \hat{\gamma}_2(y_i) + \sigma^2 +             {\cal O}(\tau^2),
\label{yM2_2}
\end{eqnarray}
\label{yM_2}
\end{subequations}
where $y(t)$ is taken in the range $y_i\pm\Delta y/2$ for
each bin $i$, and $\Delta y$ depends on the binning considered.

While the functions $\hat{m}_i$ and 
$\hat{\gamma}_i$ ($i=1,2$) are obtained explicitly for each
bin value $y_i$, the decisive aspect of this approach is the fact that the ansatz functions $m_i$ and $\gamma_i$ 
can be calculated from  the drift and diffusion coefficients $D_1$, $D_2$ and 
the measurement noise distribution $f_{\sigma}$  as follows \cite{Lind2010a}:
\begin{subequations}
\begin{eqnarray}
\gamma_1(y) &=& \int_{-\infty}^{+\infty} 
                (x-y)\bar{f}_{\sigma}(x\vert y)dx \label{gamma1}\\
            & & \cr
\gamma_2(y) &=& \int_{-\infty}^{+\infty} 
                (x-y)^2 \bar{f}_{\sigma}(x\vert y)dx \label{gamma2}\\
            & & \cr
m_1(y)      &=& \int_{-\infty}^{+\infty} 
                 D_1(x) \bar{f}_{\sigma}(x\vert y)dx \label{m1}\\
            & & \cr
m_2(y)      &=& 2\int_{-\infty}^{+\infty} 
                [(x-y)D_1(x)+D_2(x)] \bar{f}_{\sigma}(x\vert y)dx , \cr
  & & \label{m2}
\end{eqnarray}\label{gamma-m}\end{subequations}
where $\bar{f}_{\sigma}(x\vert y)$ is the probability for the system to 
adopt the value $x$ if a measured value $y$ is observed.

The problem we then  solve  is to parametrize $D_1$,$D_2$ and the noise strength $\sigma$ and to determine the set of  parameters
that minimize the functional distance $F$ between the measured functions 
$\hat{m}_i$, $\hat{\gamma}_i$ and their counterparts ${m}_i$, 
${\gamma}_i$ defined as
\begin{eqnarray}
F &=& \frac{1}{M}\sum_{i=1}^{M} \Big [ 
               \frac{\left ( \hat{\gamma}_1-\gamma_1(y_i)\right )^2}
                    {\sigma^2_{\hat{\gamma}_1}(y_i)} + 
               \frac{\left ( \hat{\gamma}_2-\gamma_2(y_i)-\sigma^2\right )^2}
                    {\sigma^2_{\hat{\gamma}_2}(y_i)} + \cr
   & & \cr
   & & \phantom{\frac{1}{M}\sum_{i=1}^{M} \Big [}
               \frac{\left ( \hat{m}_1-m_1(y_i)\right )^2}
                    {\sigma^2_{\hat{m}_1}(y_i)} +
               \frac{\left ( \hat{m}_2-m_2(y_i)\right )^2}  
                    {\sigma^2_{\hat{m}_2}(y_i)} \Big ] ,
\label{F}
\end{eqnarray}
where the summation extends over all $M$ bins, and $\sigma_{\hat{\gamma}_1}(y_i)$
is the error associated to function $\hat{\gamma}_1$ at the value
$y_i$ (and similarly for $\sigma_{\hat{\gamma}_2}$, $\sigma_{\hat{m}_1}$
and $\sigma_{\hat{m}_2}$, all of them taken directly from the data only).

After computing the functions $\hat{\gamma}_1$, $\hat{\gamma}_2$, 
we  start from the initially estimated set of values for the 
parameters and iteratively improve the solution by seeking lower values 
of $F$, until convergence is attained.  
In the following section we use two different methods for minimizing the 
functional $F$, namely the LM method \cite{Lind2010a} and SA method.
We will show that for our purposes SA is significantly better than LM.

\section{Comparing two optimization methods: Levenberg-Marquardt 
and Simulated Annealing}
\label{sec:optimization}

In this section, we consider the parameters $\sigma, d_{11}, d_{20}, d_{21}$ 
and $d_{22}$ which we denoted as $p_k$ with $k=1,\dots,5$, respectively.
For the Levenberg-Marquardt procedure one computes the  
derivatives of $F$ with respect to the parameters $\mathbf{p}= \{ p_k\}$, 
and uses them to define the coefficients 
$\beta_k = \frac{1}{2} \frac{\partial F}{\partial p_k}$
and
$\alpha_{kl} = \sum_{j=1}^4 \sum_{i=1}^{M} \frac{1}{\sigma_j^2 (y_i)} 
\frac{\partial f_j(y_i,\mathbf{p}) }{\partial p_k}\frac{\partial 
f_j(y_i,\mathbf{p}) }{\partial p_l} $, where $f_j (y_i,\mathbf{p})$ 
is the $j$-th summand in the bracketed inner sum defining $F$,  Eq.~(\ref{F}).
The descent $\delta p_l$  is computed from $\sum_{l=1}^{M} \alpha^{\prime}_{kl} 
\delta p_l = \beta_k$, which uses a damped version of the Gaussian 
matrix:
\begin{equation}
\alpha^{\prime}_{kl} = \left\{ \begin{aligned}
\alpha^{\prime}_{kl} &= (1-\lambda) \alpha_{kl} , k=l\\
\alpha^{\prime}_{kl} &=  \alpha_{kl} , k \neq l \end{aligned} \right. 
\end{equation}
The parameter
$\lambda$ is updated during the optimization. 
The procedure stops after sufficient convergence has been achieved.
For details, see Ref.~\cite{numrecip}. 
We generated several  data sets from Eq.~(\ref{xlangevin}) with different 
`input' measurement noise amplitudes $0 < \sigma_I \le 1.2$, in order to compare the reconstructed parameters to the actual values.
\begin{figure}[htb]
\begin{center}
\includegraphics*[width=14cm,angle=0]{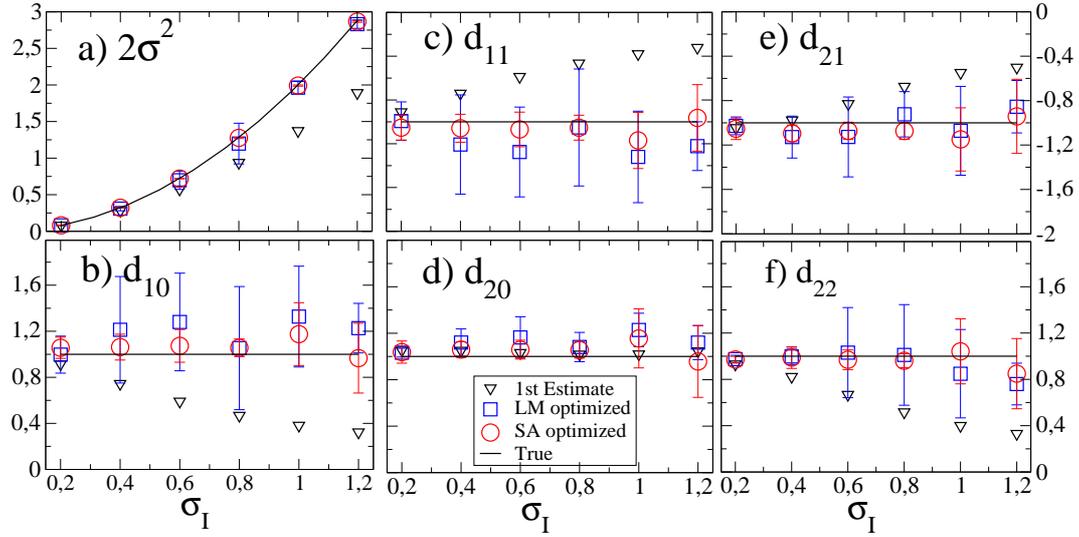}
\end{center}
\caption{\protect
        Comparison of the optimized parameters values (bullet)
        with the first estimate and the true values for different
        input measurement noise strengths $\sigma_I$:
        {\bf (a)} $2\sigma^2$,
        {\bf (b)} $d_{10}$,
        {\bf (c)} $d_{11}$,
        {\bf (d)} $d_{20}$,
        {\bf (e)} $d_{21}$,
        {\bf (f)} $d_{22}$.
        The measurement noise is correctly extracted, as well as
        the parameters representing the drift coefficient $D_1(x)$
        and the diffusion coefficient $D_2(x)$.
        SA results are significantly better than LM results (see text). Error bars represent the maximum and minimum values encountered in 10 realizations. } 
\label{fig4}
\end{figure}

The optimization results are shown in Fig.~\ref{fig4},  triangles indicating 
the first estimate for the  parameter values (see Sec.~\ref{sec:framework}),
solid lines indicating the true values used to generate the data, and squares
indicating the value after LM-optimization.

While the parameters are well estimated, particularly the magnitude of 
measurement noise and the drift parameters $d_{10}$ and $d_{11}$, it was
found that LM sometimes does not converge, or converges to a 
local minimum, the latter fact becoming visible when comparing the results 
after starting from different initial values.
Both these shortcomings required that the LM optimization results were
drawn as the best results of a series of attempts with different starting 
values. 
The intent to  overcome these shortcomings led us to attempt the SA method (squares in
Fig.~\ref{fig4}).

Simulated annealing is a  probabilistic  global optimization method.  
Starting  from initial guesses for the parameter vector it proceeds with a 
step in a  random direction. The step is accepted immediately if the energy 
function at the new position $F_{\mathrm{new}}$ (see Eq.~(\ref{F})) 
is lower than the previous 
energy $F_{\mathrm{old}}$, or it is accepted with a probability given by a 
Boltzmann factor $\exp(-\Delta F/kT)$, where $\Delta F = F_{\mathrm{new}} - 
F_{\mathrm{old}}$. Otherwise the step is refused and a new parameter step is 
chosen. This procedure corresponds to the motion of thermal atoms in an 
attractive  potential. Gradual cooling is then achieved by reducing the 
temperature parameter, thus annealing the test particle in the minimum of 
the energy landscape.

Figure \ref{fig4} shows that both  optimization methods determine the value 
of input measurement noise $\sigma_I$  correctly in all cases, which we find remarkable given the 
comparable magnitude of measurement noise and the time series $x(t)$ for 
$\sigma \approx 1$. 
Furthermore, it can be seen that both LM and SA estimate the parameters 
$d_{1k}$ for the drift and $d_{2l}$ for the diffusion coefficient correctly; 
as expected \cite{Lind2010a}, the drift coefficients are determined with higher 
precision. 
In general, the SA routine provides more accurate results than LM.

In the LM case, for stronger measurement noise amplitudes, 
namely for $\sigma>1.2$,  
the algorithm is sometimes stuck in a local minimum of the function $F$, 
leading to unreliable coefficients $d_{ik}$. 
The same set of $10$ time series with $10^6$ data points were used
for each method,
but in the LM case, not all cases showed local minima sufficiently
close to the true parameter values. Therefore, in our comparison,
we overestimate the best performance of the LM method, by considering 
only the best five cases out of the set of $10$ time series
(squares in Fig.~\ref{fig4}).
\begin{figure}[htb]
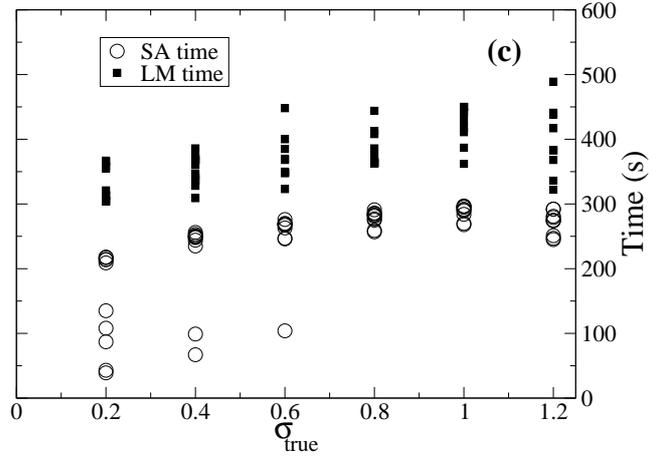

\begin{center}
\includegraphics*[height=6cm,angle=0, bb=0 40 440 410, clip]{fig05a_joao.eps}%
\includegraphics*[height=6cm,angle=0]{fig05b_joao.eps}%
\end{center}
\caption{\protect
        {\bf (a)} Plot of the relative energy in comparison to the true value, Eq.(\ref{dF}), as a function of the Euclidean distance 
        $dr=||p_i-p_{i,0}||_2$
        of the parameters from the true parameter values 
        $\{ p_{i,0} \}$ found with
        SA method (circles) and with the LM method (squares) for $\sigma=1.2$. 
        {\bf (b)} Close-up of (a) near the minimum of $F$. 
        For better visibility, not all SA-points are shown in (a) 
        and (b).
        {\bf (c)} Execution time (in seconds) as a function of $\sigma$
        for LM (squares) and SA (circles).}
\label{fig5}
\end{figure}

Figure \ref{fig5}a shows a plot of the energy, i.e.~the distance  
\begin{equation}
dF=F(p_i)-F_0(p_{i,0})
\label{dF}
\end{equation}
where $F_0(p_{i,0})$ is the value found when taking the true
parameter values, as a function of the Euclidean distance $dr=||p_i-p_{i,0}||_2$ 
of the parameters from the true 
parameters values. 
Apparently, the energy is strongly asymmetrical in at least one of the 
parameters, which is in agreement with Ref.~\cite{Lind2010a}. 
Also, the existence of local minima,  which apparently attract the LM 
algorithm, is confirmed.
Figure \ref{fig5}b shows a close-up near the lowest minimum to emphasize 
the higher accuracy of the final result for SA (circles) compared with LM 
(squares). In this example, it can be seen that the LM algorithm does not 
converge to the global minimum but deviates shortly before.

Another argument in favor of the SA algorithm is that the execution time
approximately $100$ s lower than for the LM method, as shown in
Fig.~\ref{fig5}c. Additionally, whereas 
the LM algorithm needs to  execute on $10$ realizations
of the time series in 
order to produce a 
reasonably accurate result, the SA algorithm was found to converge to 
the global minimum in almost all instances, implying that fewer 
realizations are needed for averaging.

\section{Discussion and Conclusions}
\label{sec:conc}

We describe  the improvement of a  nonparametric procedure to extract measurement 
noise in empirical stochastic series with strong measurement noise by applying a simulated annealing approach for the optimization step.
Simulated annealing accurately extracts the strength of measurement 
noise and the values of the parameters representing drift and diffusion. These parameters
fully describe the evolution equation for the measured quantity. SA produces more reliable and accurate results than a previously used Levenberg-Marquardt algorithm, with the additional benefit of being significantly faster.

Nonparametric reconstruction of the
Langevin Eq.~(\ref{xlangevin}) from measured stationary data sets
merely requires that the process exhibits Markovian properties and 
fulfills the Pawula theorem \cite{lind05}, whereas the measurement noise needs to be uncorrelated. 
The Pawula constraint can be relaxed extending the analysis to L\'{e}vy 
noise \cite{friedrich08,siegert01}. It should be pointed out  that the method 
presented here  relies solely on the Markovian properties of the underlying, 
undisturbed process $x(t)$, and does not require the measured process $y(t)$ to be Markovian.

An extension of this work to multidimensional Langevin time series  will be essential for assessing  the complexity behind EEG time series or earthquake data. It is likely that we may combine this denoising approach with the method of eigendirections presented in \cite{vitor}.

\section*{Acknowledgments}

The authors thank D.~Kleinhans, M.~W\"achter, J.~Peinke and B.~Lehle
for useful discussions.
They also wish to thank {\it Funda\c{c}\~ao para a Ci\^encia e
a Tecnologia} (FCT) for the {\it Ci\^encia 2007} support (PGL), for the fellowship
SFRH/BPD/65427/2009 (FR), and for the BII student support (JC). To FCT and {\it Deutscher Akademischer Austauschdienst} (DAAD) the authors express their gratitude for support through the bilateral cooperation DREBM/DAAD/03/2009 between Portugal and Germany. Finally, they are indebted to the  organizers and participants of Dynamics Days 2010 South America for useful discussions in an inspiring atmosphere.


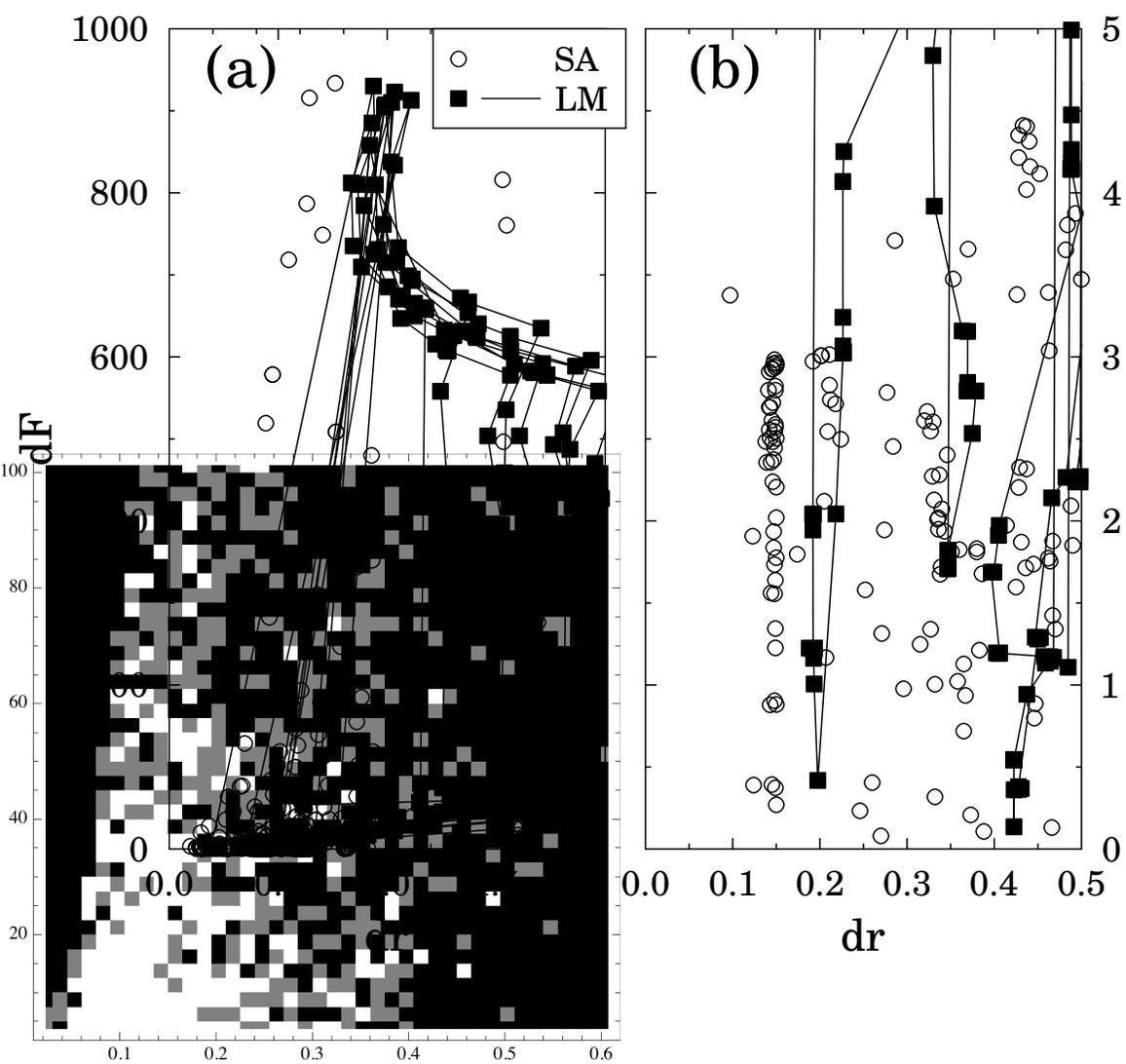
\end{document}